\documentstyle[preprint,aps]{revtex}
\begin{document}
\draft
\title{THE CHARGED NEUTRINO: A NEW APPROACH TO THE SOLAR
NEUTRINO PROBLEM}
\author{A.Yu.Ignatiev\cite{byline1} and
G.C.Joshi\cite{byline2}}
\address{Research Centre for High Energy Physics, School of
Physics, University of Melbourne, Parkville, 3052, Victoria,
Australia}
\date{March 22, 1994}
\maketitle
\begin{abstract}
We have considered the effect of the reduction of the solar
neutrino flux on earth due to the deflection of the charged
neutrino by the magnetic field of the solar convective zone.
The antisymmetry of this magnetic field about the plane of
the solar equator induces the anisotropy of the solar
neutrino flux thus creating the deficit of the neutrino flux
on the earth. The deficit has been estimated in terms of
solar and neutrino parameters and the condition of a 50 \%
deficit has been obtained: $Q_{\nu} gradH \geq 10^{-18}
eG/cm$
where $Q_{\nu}$ is the neutrino electric charge, $gradH$ is
the gradient of the solar toroidal magnetic field, e is the
electron charge. Some attractive experimental consequences
of this scenario are qualitatively discussed.
\end{abstract}

The discrepancy between the results of the solar neutrino
experiments \cite{home,kamioka,sage,gallex} and the
predictions of the Standard Solar Models \cite{ssm1,ssm2} is
one of the most challenging issues of modern particle
physics and astrophysics.
Some of the plausible solutions to the solar neutrino
problem are neutrino oscillations amplified by the Mikheev-
Smirnov-
Wolfenstein effect \cite{MSW,wolf}, neutrino decay
\cite{decay}, neutrino spin-precession \cite{OVV} in the
solar magnetic field and resonant spin-flavour conversion
scenario \cite{RSF} (assuming the twisting structure of the
solar magnetic field in the convective zone has recently led
to a more complicated variant of the last scenario \cite
{twist}).
However, all the solutions proposed so far rely on some
hypothetical properties of the neutrino and/or the solar
magnetic field which have to be confirmed by future
evidence--hence no final solution emerged as yet.

In this paper we suggest a quite different approach to the
solar
neutrino problem which does not employ any detailed
assumptions about the small scale structure of the solar
magnetic field or the variation of solar density. We assume
that the electric charge of the neutrino is non-zero. As for
the solar magnetic field, the only property we use is--a
well-established fact--its antisymmetry with respect to the
plane of the solar equator. In this way, our approach
belongs entirely to the realm of classical physics and the
Lorentz force is the only essential theoretical tool we
need.

In fact, the neutrino is the only
elementary particle, besides the gauge bosons, whose
electric
charge is normally assumed to be zero. But if the neutrality
of the gauge bosons is deeply rooted in the principle of
gauge invariance, there are no compelling reasons whatsoever
for the neutrino to have zero charge.

Of course, the neutrino is assumed to be exactly neutral
within the Standard Model. However, the
recently  developed
approach to the problem of the
electric charge quantization has led to the realization of
the fact that in a fairly large class of gauge models,
including the  Minimal Standard Model, the electric charge
can be
dequantized \cite{Mel} (see also \cite{I}). This
means that
the electric charges of elementary particles can take
different values from those conventionally assumed:
 $Q_{\nu}=0$, $Q_{l}=-e, Q_{u,c,b}=2e/3$ and $Q_{d,s,b}=-
e/3$ (e being the modulus of the electronic charge).
In
particular, the neutrino can acquire nonzero electric
charge. (Another interesting aspect of the theories with
dequantized
electric charges is that one might speculate about the
possibility of time dependence of the electric charges
within such theories \cite{we}.)

In other words, in Refs. \cite{Mel} it has been shown that
the Standard Model contains an additional
free parameter $\epsilon$ which must be determined
experimentally along with the other more familiar parameters
such as Higgs mass or Yukawa couplings. Of course if it were
found that $\epsilon$ is nonzero, it should be very small
anyway (see discussion below) and that would create one more
hierarchy
problem. Yet taking into account the existence of a few such
problems already, the appearance of a new one does not seem
strong enough argument to disregard the possibility of
nonzero $\epsilon$.

Furthermore, one might argue that nonzero neutrino charge
does not follow from any theoretical principle, whether
established or hypothetical (with the exception of the well-
known rule "all which is not forbidden is allowed"). But
now, based on the works \cite{Mel}, we know that the zero
neutrino charge does not follow from anywhere, too!

Another possible objection against particles with small
fractional charge is that it is difficult to embed them into
grand unified theories \cite{OVZ}. Yet theories with
paraphoton provide a viable alternative \cite{holdom}.

So, at present the cases of zero/nonzero neutrino charges
must be considered as two working hypotheses on the equal
footing, only experiment being capable to provide the
ultimate answer. The situation with neutrino charge is very
similar to the situation with neutrino mass: while zero mass
is the prediction of the minimal standard model, most
physicists agree that the question of zero/nonzero neutrino
mass has much more to do with experimenting than with model-
building. While it would not be easy to detect the neutrino
charge, the consequences of such discovery should certainly
be dramatic, ranging from the prospects of detecting relic
neutrino through its electromagnetic interaction to possible
better ways of managing neutrino beams, creation of neutrino
optics etc.

Finally, let us note that in the present work we are not
concerned if the neutrino mass is zero or not. Certainly,
there exist well-known difficulties associated with
charged massless particles \cite{massless}. However, one can
take a pragmatic point of view \cite{brf} and keep
developing a theory until one runs into any inconsistency.
No such inconsistency seems to show up in our treatment.
An alternative point of view is to give the neutrino a Dirac
mass by introducing additional Higgs multiplets.

Note also that even if the neutrino is massless in vacuum,
it cannot be considered massless inside plasma. This is
because the vacuum dispersion relation $E=| {\bf p}|$ gets
changed by the weak interaction of neutrino with plasma
\cite{wolf}. In other words, there arises a refraction index
for the neutrino propagating through plasma. Thus, the
situation with infrared divergencies might be better for a
neutrino in plasma than in vacuum.

To conclude, it seems that assuming nonzero neutrino charge
is certainly not more heresy than assuming nonzero neutrino
magnetic moment, or mass and mixing angles.

In addition, there exist quite an independent motivation to
study the behaviour of a charged neutrino inside the Sun.
The point is that the neutrino electromagnetic properties
get modified by plasma effects and under certain conditions
these modifications result in {\em an induced electric
charge of the neutrino\/} \cite{induced}. We stress that it
happens in the Minimal Standard Model where the neutrino has
zero {\em intrinsic\/} electric charge.

The purpose of  this paper  is to draw attention to the fact
that the possible existence of a very small  electric charge
of the neutrino may be a clue to the solar neutrino
problem\footnote{The possible role of the charged neutrino
interaction with the terrestrial electric field, in
connection with the solar neutrino problem, was discussed
previously by G.C.Joshi and R.R.Volkas (unpublished).}. The
idea is that the charged  neutrinos  are
deflected by the solar  magnetic field  while passing
inside the  Sun. Due to the antisymmetry of that magnetic
field about the solar equatorial plane the resulting
neutrino flux is made
anisotropic which leads to the solar neutrino deficit
registered on the Earth. (Within our scenario, this deficit
is not real but only apparent in the sense that {\em the
total \/} $4 \pi$ solar neutrino flux is {\em not changed\/}
as compared with the standard solar models.)

Our key result is:
\begin{equation}
{\Phi_1 \over \Phi_0} \equiv 1+ \delta_0= 1+ {\epsilon e
\langle
{\partial H_{\phi} \over \partial z} \rangle d D \over E} ,
\label{12}
\end{equation}
where (all units are Gaussian) $\Phi_1$ is the neutrino flux
observed on the
Earth, $\Phi_0$ is the flux predicted by the standard
solar model, $Q_{\nu}= \epsilon e$ is the neutrino electric
charge,  $ \langle {\partial H_{\phi} \over \partial z}
\rangle$ is the average gradient of the toroidal magnetic
field in the solar convective zone, taken along the $z$-
axis, parallel to the rotation axis of the Sun, $D$ is the
width of the convective zone, $d$ is the distance from the
solar centre to the middle of the convective zone, E is the
neutrino energy. The detailed derivation and discussion of
this formula are to be given elsewhere \cite{preprint}.
Here, we note that the main assumption behind the
Eq.~(\ref{12}) is the smallness of the ratio  $
\delta_0=\epsilon e \langle
{\partial H_{\phi} \over \partial z} \rangle d D/ E$
compared to unity (more exactly, it has to be $\delta_0 \alt
0.7$).

To obtain, say, a 50 \% deficit, one needs to have
\begin{equation}
| \epsilon \langle {\partial H_{\phi} \over \partial z}
\rangle | \agt 10^{-18} G/cm  \label{13}
\end{equation}
(assuming $D= 2 \times 10^{10}$ cm, $d= 6 \times 10^{10}$
cm, and $E= 0.8$ {\rm ~MeV}).

 Since the
magnetic field reverses itself with a period of 11 years,
our Eq.~(\ref{12}) implies that each 11 year period of
neutrino flux {\em deficiency\/} must be followed by 11 year
period of neutrino flux {\em excess of the same magnitude\/}
so that the flux averaged over the 22 year cycle would be
the same as predicted by the Standard Solar Model.

There are several possibilities to overcome this difficulty.

The most natural one is to go beyond linear approximation on
which Eq.~(\ref{12}) is based. This would be definitely
required if $| \epsilon \langle {\partial H_{\phi} \over
\partial z} \rangle | \agt 2 \times 10^{-18}$ G/cm.

Naively, one might expect that the neutrino deficiency must
alternate with the neutrino excess at 11 year intervals
independently of the magnitude of the gradient: just note
that  when the magnetic field configuration is defocusing,
one would expect the neutrino deficiency and when it is
focusing, the neutrino excess. Each reversal of the magnetic
field
means a switch between focusing and defocusing modes so that
any 11 year "deficiency" cycle would be followed by the 11
year
"excess" cycle, however great the gradient of the magnetic
field is. Nevertheless, there are arguments based on simple
geometrical optics considerations which show that it is not
the case
and if the gradient is large enough then the neutrino
deficiency can occur both for the defocusing {\em and the
focusing\/} configuration!

Another option is to try to relax the solar upper bound on
the possible electric charge of the electron neutrino
obtained in \cite{brf}, since the neutrino charge and the
magnetic field gradient come always as the product $\epsilon
\times \langle {\partial H_{\phi} \over \partial z} \rangle
$
rather than separately.

Let us also mention briefly that at present we cannot rule
out the possible existence of a primordial magnetic field of
as much as $10^{6}$ G inside the core of the Sun
\cite{parker}. Within the present context, it would very
interesting if any evidence could be obtained concerning the
existence of significant gradients of that field near the
 plane of the solar equator.

Finally, although it is not as much appealing, we should not
discard the possibility that our mechanism is effective only
during alternative 11 cycles or even only during the periods
of active sun within the alternative 11 year cycles while
some other mechanism is responsible for neutrino depletion
during the rest of the time. This possibility will have to
be considered much more seriously if the anticorrelation of
the neutrino deficiency with solar activity is established
firmly by the future experiments.

Now, assuming that the above default is cured in one or
another way, let us turn to the experimental implications of
our result. At this stage, the status of both the present
scenario and the experimental data does not encourage one to
make detailed quantitative comparison of theory and
experiment. Rather, we confine ourselves to a qualitative
attempt to match the general consequences of the proposed
hypothesis with the outstanding features of the available
data. In this way, one can easily see that our result,
Eq.~(\ref{12}), does point to the right direction while
confronted with the following main experimental conclusions:

1) Anticorrelation of the neutrino flux with solar activity
is probably observed in the Homestake data \cite{anti,bapr}.

2)No such anticorrelation is observed in the Kamiokande data
\cite{kamioka}.

3)The higher neutrino flux (i.e., less neutrino deficit) is
observed in Kamiokande experiment than in Homestake
experiment.

4)The higher neutrino flux is observed in SAGE \cite{sage}
and GALLEX \cite{gallex} experiments than in Homestake
experiment.

The reason is that the experimental
thresholds of neutrino energy are rather different in those
experiments: $E_{Home}=0.816$ {\rm ~MeV}, $E_{Kam} \sim 7.5$
{\rm ~MeV}, and $E_{Gallex}=0.233$ {\rm ~MeV}, while our
result, Eq.~(\ref{12}) scales in  the ratio
$\epsilon \langle {\partial H_x \over \partial z} \rangle
/E$. Therefore, changing the magnetic field will be
equivalent to changing the neutrino energy
correspondingly.
Furthermore, it is natural to assume that both neutrino flux
deficit and anticorrelations grow with the increase of the
gradient. Hence we obtain that the anticorrelations have to
be {\em smaller\/} for {\em more energetic \/} neutrinos.
And this is exactly what is needed to qualitatively explain
the difference between Homestake and Kamiokande data (see 1)
and 2) above). Also, by the same reasoning, within our
scenario one can expect less deficit in Kamiokande than in
the Homestake experiment.

Now, as for the fourth feature, i.e., results of gallium
experiments, our hypothesis seems to predict {\em greater\/}
deficit than Homestake and thus looks disfavored by gallium
results. However, one must remember that: 1) the difference
between Gallium and Homestake results, from the viewpoint of
our hypothesis, must be {\em less\/} pronounced than the
difference between Homestake  and Kamiokande data. This
follows from the fact that the ratio of the characteristic
neutrino momenta for Homestake-Gallium data are, roughly,
less than for Kamiokande-Homestake data by a factor of 3:

\begin{equation}
{E_{Kam} \over E_{Home}} \simeq 10,{E_{Home} \over E_{Ga}}
\simeq 3;
\end{equation}
2) The errors of Gallium data are still larger than those of
Homestake data.

Now, we would like to draw attention to a curious
coincidence in the solar neutrino data. Kamiokande does not
see anticorrelations during the whole period of its
operation, i.e., 1987-1993  (part of
solar cycle \# 22). And, according to \cite{bapr}
there are no anticorrelation in  Homestake data during the
years 1970-1977 (of which the period 1970-1976 is a part of
solar cycle \# 20). Also, the
latest data do not confirm the anticorrelation: large number
of the sunspots in 1991--1992 was accompanied by high
counting rate \cite{smirnov}. Therefore, one is tempted to
speculate
that, due to some reason, the anticorrelations are much more
prominent in the {\em odd-numbered\/}  solar cycles while
being
suppressed in the {\em even-numbered\/}  cycles. If we take
this
conjecture seriously, it would be easy to conclude that the
neutrino-depleting mechanism must somehow be correlated not
only with {\em the strength\/} of the solar magnetic field
but also with {\em the direction\/} of the toroidal solar
magnetic field which reverses every 11 years. Obviously,
this
feature would be difficult to accomodate within any of the
existing scenarios  except the present one.

Apart from the reduction of the conventional
(i.e., thermonuclear) neutrino flux, a spectacular feature
of
our scenario is the prediction of a "second flux" of
electron neutrinos and antineutrinos from the Sun. While
thermonuclear neutrinos are produced due to the {\em weak\/}
interactions of the neutrino, the second flux arises due to
the {\em electromagnetic\/} production of neutrino-
antineutrino pairs. The most important process would be that
of plasmon decay into a neutrino-antineutrino pair. Thus the
second flux would consist of low-energy (about 200 {\rm eV})
 neutrinos produced in plasmon decays in the core of the
Sun, the number of such neutrinos being much greater than
that of the thermonuclear neutrinos. It would be
very interesting to consider the posibility of detecting
this second neutrino flux, about $ 10^{16} \times
(\epsilon/10^{-13})^2$ $s^{-1}cm^{-2}$
in magnitude, on the Earth.

 Let us now turn to the current limits on the neutrino
electric charge and the gradient of the solar magnetic field
to assess if the criterion Eq.~(\ref{13}) can realize or
not.

Various bounds on the
neutrino charges have recently been analysed in a systematic
way in Ref. \cite{bv}. The strongest model-
independent\footnote{By model-
independent we mean the constraints that do not rely on
additional assumptions such as charge conservation or the
equality $Q(\nu_{e})=Q(\bar{\nu_{e}})$.} constraints on the
electron
neutrino charge, $Q(\nu_{e})= \epsilon e$, come from three
sources: analysis of $\nu_{e} e$ elastic scattering
\cite{brf}: $\epsilon \alt 3 \times 10^{-10}$; study of
plasmon decay into neutrino-antineutrino pairs
\cite{brf}: $\epsilon \alt 10^{-13}$; detection of a
neutrino
signal from SN1987A supernova explosion \cite{bc}:
$\epsilon \alt 10^{-(15 \div 17)}$.

Yet it is generally believed (see, e.g.\cite{partdata}) that
the constraint based on SN1987A arguments, although
stronger, is less reliable than the previous ones because it
involves the
details of the galactic magnetic field which are not very
well known.

There exist even more severe, but less direct constraints.
They  are based on the experimental data on the neutrality
of
atoms and neutrality of the neutron. These data give limits
on the sum of the proton and electron charges \cite{ep}:
$Q(p)+Q(e)= (0.8 \pm 0.8) \times 10^{-21}e$ and the neutron
charge \cite{n}: $Q(n)= (-0.4 \pm 1.1) \times 10^{-21}e$.
Then, assuming charge conservation in the neutron beta-decay
$n \rightarrow p+e^{-}+ \bar{\nu_{e}}$  we can obtain the
bound on the electron {\em anti}neutrino charge: $ Q(
\bar{\nu_{e}}) < 3 \times 10^{-21}e$. Finally, assuming
validity of CPT symmetry\footnote{Besides charge
conservation and CPT,  a number of usually unspoken but very
important assumptions underlying the last constraint
are made. For instance, one has to assume that the electric
charges of {\em free\/} electrons and protons are exactly
the same as those of {\em atom-bound\/} electrons and
protons. Another fundamental assumption, as noted in Ref.
\cite{brf}, is that the electric charge, as measured by
interaction with an electromagnetic field, coincides with
the electric charge assigned by the charge-conservation law
(see also a discussion of that point in \cite{fg}).
According to Ref. \cite{brf}, it is possible to construct
models in which it is not the case.  Under ordinary
circumstances there is no doubt in the correctness of the
above axioms, but when it comes to such outstanding
accuracies
as $10^{-21}$, it does not seem unreasonable to question
those axioms, too.} with respect to $\nu_{e}$ and $
\bar{\nu_{e}}$ charges, one can claim that $
 Q(\nu_{e}) < 3 \times 10^{-21}e $. Yet the requirements of
the electric charge
conservation and CPT symmetry, although very general and
perfectly valid up to now, are themselves a subject of
current
experimental testing \footnote{Note that it is
possible to constrain the $\nu_{e}$ charge assuming only
charge conservation in the decay $ \beta^+ $ decay, but not
the equality
$Q(\nu_{e})=Q(\bar{\nu_{e}})$. Naturally, this constraint
turns out to be much weaker than $3 \times 10^{-21}e$
namely:
$Q(\nu_{e}) < 4 \times 10^{-8}e$ \cite{hughes}.}.
Furthermore, there exist
several models in which charged neutrino arises as a natural
consequence of the electric charge violation \cite{Qnoncons}

Note also that recently there has been considerable interest
in discussing the possible existence of new particles
carrying very small electric charge  ("milli-charged
particles") \cite{I}. These works contain detailed
discussion of many phenomenological constraints on such
particles obtained from a variety of sources (including
astrophysics, cosmology, geophysics and macroscopic
electrodynamics). Many of those constraints apply to the
case of electron neutrino, too; we shall not repeat that
material here.

Now, let us discuss the gradient of the toroidal magnetic
field in the convective zone of the Sun.
A crude estimate of the
gradient can be obtained by dividing $H$ by $h$ where $H$ is
the maximum value of the magnetic field reached at the
latitudes of about $ \pm 10^o$ \cite{parker} and $h$ is the
distance from that latitude to the solar equatorial plane,
$h=d \sin{10^o} \approx 10^{10}$cm.

As for the possible
value of $H$, it is a subject of a debated controversy.  On
the one hand, it is claimed \cite{smallH} that values of $H$
greater than $10^{4} G$ are ruled out by the non-linear
growth-limiting effects; on the other hand, there are
arguments based on the helioseismology data that it can
reach as large values as a few million G \cite{largeH}.
Anyway, magnetic fields up to $10^{4}$ G (or even $10^{5}$
G, see e.g.\cite{akh}) are widely used by many authors
trying to
explain the solar neutrino puzzle. So, we leave it to the
reader to make his/her own judgement on that point. Note
also, that it is the magnetic field close to the surface of
the Sun which reaches its maximum at $10^o$ latitude, and
this latitude may be higher (or lower) for magnetic fields
located at larger depths. That brings in an additional
uncertainty to the estimate of the gradient. If we do admit
that the magnetic field in the convective zone may vary in
the range $H= 10^{3} \div 10^{6}G $ than the value of the
gradient may vary in the range

\begin{equation}
\langle {\partial H_{\phi} \over \partial z} \rangle  \simeq
{H_{\phi}
\over h} \approx (10^{-7} \div 10^{-4}) \; G/cm .\label{131}
\end{equation}
Hence we see that if we take $\epsilon \simeq 10^{-13}$ as a
conservative upper bound on the neutrino charge, the value
of the gradient needed  to explain
the neutrino deficit, $ \langle {\partial H_{\phi} \over
\partial z} \rangle \simeq 10^{-5}$ G/cm {\em may indeed
exist
in the convective zone of the Sun\/}.

To conclude, in the context of the solar neutrino problem we
studied the consequences of the hypothesis that the electron
neutrino has a small but non-vanishing electric charge. The
main general consequence is that the solar neutrino flux can
be anisotropic. That anisotropy is driven by the Lorentz
force acting on the charged neutrino on the part of the
solar toroidal magnetic field which is antisymmetric
 about the solar equatorial plane. The general formula for
the neutrino flux deficit is obtained  which leads to a
certain condition on the product of the neutrino electric
charge and the gradient of the magnetic field which has to
be met to obtain an observed value of the deficit.

We then discussed some attractive experimental implications
of this scenario as well as the problems which have to be
solved so that this scenario could be considered as a full-
fledged solution to the solar neutrino puzzle.

Independently of whether this scenario survives or not in
its
present form, our arguments show that a more general problem
of the possible anisotropy of the neutrino flux due to the
interactions of the neutrino with the solar matter and
electromagnetic fields is certainly worth further pursuing.

The authors are grateful to N.Frankel, V.Gudkov, A.Klein,
B.McKellar, and R.Volkas for fruitful discussions. A.I. is
indebted to A.Dymnikov for valuable help. This work was
supported in part by the Australian Research Council.

\end{document}